\documentclass[a4paper,11pt]{article}
\usepackage{amsmath,amsfonts,amssymb,amsthm,eucal}
\usepackage[T1]{fontenc}
\usepackage[utf8]{inputenc}
\numberwithin{equation}{section}

\newtheorem{lem}{Lemma}[section]
\newtheorem{thm}[lem]{Theorem}
\newtheorem{cor}[lem]{Corollary}
\theoremstyle{remark}
\newtheorem{rem}[lem]{Remark}
\newtheorem{prop}[lem]{Proposition}
\theoremstyle{definition}
\newtheorem{defn}[lem]{Definition}
\theoremstyle{remark}

\begin{document}

\title{A Transmission Framework for Stark Operators with
$\delta$ Interactions on a Compact Lipschitz Interface}
%\author{Masahiro Kaminaga}
\author{%
Masahiro Kaminaga\thanks{Corresponding author.
Email: kaminaga@mail.tohoku-gakuin.ac.jp}\\
{\small Department of \textit{Information Technology}, Faculty of Engineering,}\\
{\small Tohoku Gakuin University, Sendai, Japan}\\
{\small ORCID:\ 0000-0001-7204-8300}%
}
\date{}
\maketitle



\begin{abstract}
We study an elliptic transmission problem associated with a Stark operator and a $\delta$ interaction on a compact Lipschitz interface in $\mathbb R^d$.
The background differential expression contains the unbounded coefficient $-Fx_1$, and the interface strength is an arbitrary real function $\alpha\in L^\infty(\Sigma)$.
We introduce a transmission class with piecewise $H^1$ regularity near the interface and an $L^2$ action away from it.
This setting gives Dirichlet traces in $H^{1/2}(\Sigma)$ and weak normal derivatives in $H^{-1/2}(\Sigma)$ without assuming smoothness of $\Sigma$.
We prove that the transmission conditions define a self--adjoint realization of the formal operator $H_{F,0}+\alpha\delta_\Sigma$.
We also obtain a boundary resolvent formula in terms of the free Stark resolvent and a bounded operator from $H^{-1/2}(\Sigma)$ to $H^{1/2}(\Sigma)$.
The formula implies that the resolvent difference is compact on $L^2(\mathbb R^d)$.
Consequently, if $F\ne0$, the essential spectrum of the interacting operator is $\mathbb R$.
The result supplies a direct boundary reduction for an interface problem whose background potential is not bounded and is not translation invariant.
\end{abstract}

\section{Introduction}

Interface problems for elliptic operators arise when a thin layer is replaced by a transmission condition on a hypersurface.
In the present paper, the bulk operator is the Stark operator and the interface carries a real bounded $\delta$ interaction.
The main difficulty is that the bulk potential is unbounded, while the interface is only assumed to be Lipschitz.
Our purpose is to give a direct transmission framework in which the operator realization, the boundary reduction, and the essential spectrum follow from one construction.

The free Stark Hamiltonian
$$
H_{F,0}=-\Delta-Fx_1,\qquad F\in\mathbb R,
$$
acting in $L^2(\mathbb R^d)$, is self--adjoint and, for $F\ne0$, has purely absolutely continuous spectrum; see Avron--Herbst \cite{AvronHerbst1977}.
Related spectral, decay, and resonance results for Stark operators were obtained in \cite{BentoselaEtAl1983,Herbst1979,Herbst1980}.
In particular,
$$
\sigma(H_{F,0})=\mathbb R,
$$
and $H_{F,0}$ has no eigenvalues when $F\ne0$ \cite{AvronHerbst1977}.

Singular interactions supported on hypersurfaces form a standard class of Schr\"odinger operators; see \cite{AGHH}.
Boundary integral methods and Kre\u{\i}n resolvent formulas for $\delta$ interactions were developed in \cite{BEKS1994}, and further results for $\delta$ and $\delta'$ interactions on smooth compact hypersurfaces were obtained in \cite{BehrndtLangerLotoreichik2013}.
Abstract Kre\u{\i}n--type formulas for singular perturbations of
self--adjoint operators were developed, for example, in
\cite{Posilicano2001}.
Transmission realizations and essential spectra for interactions on unbounded smooth hypersurfaces were studied by Rabinovich \cite{Rabinovich2019}.
Magnetic Schr\"odinger operators with $\delta$ type potentials were considered in \cite{Rabinovich2021}.
For the special case of several concentric spherical $\delta$--shell interactions, a boundary integral resolvent representation was obtained in \cite{KaminagaAAMP}.
Models with interactions supported on spheres and related penetrable--wall models were studied in \cite{AntoineGesztesyShabani1987,HounkonnouHounkpeShabani1997,IkebeShimada1991,Shabani1988JMP,Shabani1988NCB,Shimada1992,Shimada1994}.
In particular, the spectral properties of penetrable wall Hamiltonians were analyzed in the classical work of Ikebe and Shimada \cite{IkebeShimada1991}.

A physical reason for the model is the following.
The term $-Fx_1$ describes a particle in a constant electric field, while a hypersurface interaction is an ideal model for a thin material interface, a surface defect, or a narrow layer whose thickness is small compared with the relevant length scale.
The electric field permits escape in one direction, and therefore the spectral set of the free operator is already the whole real line.
This fact does not make the surface interaction trivial.
The interaction changes the resolvent, the boundary response, and in suitable extensions of the present setting the scattering and resonance data.
It is therefore useful to have a rigorous boundary reduction which keeps the Stark background and the surface interaction in one formula.

There is also a mathematical reason to separate the resolvent problem from the
location of the spectrum as a set.
The equality $\sigma(H_{F,0})=\mathbb R$ contains no information about how a
codimension--one singular perturbation changes the resolvent or the boundary
response.
For a compact interface one expects the perturbation to be relatively compact
at the resolvent level, but this is not automatic from the usual bounded
potential theory: the perturbation is distributional and the background
coefficient is unbounded.
The boundary formula below proves this compactness directly and identifies the
operator on $\Sigma$ which contains the effect of the interface.

The papers cited above show that transmission conditions and boundary operators are effective for singular interface problems.
The present setting differs in three connected points.
First, the coefficient $-Fx_1$ is unbounded in the whole space.
Second, we allow a compact Lipschitz interface and a variable strength $\alpha\in L^\infty(\Sigma;\mathbb R)$.
Third, we require a construction that is compatible with the graph domain of the Stark operator although this domain is not used as a global $H^1$ space.
These points prevent a direct use of the usual translation invariant layer potential for the Laplacian.
They also require a careful distinction between the graph trace of the free operator and the transmission trace of a single layer potential.
The algebraic form of the boundary resolvent formula is consistent with the
general theory of singular perturbations.
The analytic point specific to the present setting is to identify this
abstract structure with an explicit transmission operator.
For this purpose, we prove that the trace is bounded on the graph domain of
the free Stark operator, that the layer potential defined through the extended
Stark resolvent has the required one--sided regularity on a Lipschitz
interface, and that its weak normal derivatives satisfy the correct jump
relation.
These facts are needed because the free graph domain is not used as a global
$H^1$ space and the background coefficient is unbounded.

We consider the formal Hamiltonian
$$
H_{F,\alpha}=H_{F,0}+\alpha\delta_\Sigma,
$$
where $\Sigma$ is the boundary of a bounded Lipschitz domain.
We define $H_{F,\alpha}$ as a self--adjoint realization of $H_{F,0}$ on $\mathbb R^d\setminus\Sigma$ with transmission conditions on $\Sigma$.

Our first contribution is a transmission class that gives the weak traces needed for the interface condition without imposing global $H^1$ regularity on the Stark graph domain.
Our second contribution is a proof that these conditions give a self--adjoint realization for every real bounded interface strength.
Our third contribution is a boundary resolvent formula expressing the resolvent of $H_{F,\alpha}$ in terms of the free Stark resolvent
$$
R_{F,0}(z)=(H_{F,0}-z)^{-1}
$$
and the boundary operator
$$
M_F(z)=\tau R_{F,0}(z)\tau^*.
$$
The formula reduces the interface problem to the bounded boundary operator $I+\alpha M_F(z)$ on $H^{-1/2}(\Sigma)$.
It also proves that the resolvent difference
$$
(H_{F,\alpha}-z)^{-1}-(H_{F,0}-z)^{-1}
$$
is compact for every $z\in\mathbb C\setminus\mathbb R$.
As a consequence, when $F\ne0$,
$$
\sigma_{\mathrm{ess}}(H_{F,\alpha})=\mathbb R.
$$

Our main results are summarized below.

Throughout the paper, for a function defined on both sides of $\Sigma$, we use the notation
$$
[u]:=u^+-u^-,
\qquad
[\partial_\nu u]:=(\partial_\nu u)^+-(\partial_\nu u)^-
$$
for the jumps across $\Sigma$, where the traces are taken from the bounded and exterior regions and the same fixed normal vector $\nu$ is used on both sides.

\begin{itemize}
\item[(i)] For the boundary $\Sigma$ of a bounded Lipschitz domain, the operator $H_{F,\alpha}$ admits a self--adjoint realization with transmission conditions
$$
[u]=0,
\qquad
[\partial_\nu u]=\alpha\gamma u.
$$

\item[(ii)] For $z\in\mathbb C\setminus\mathbb R$, the resolvent of $H_{F,\alpha}$ admits a boundary representation in terms of the free Stark resolvent and the operator
$$
M_F(z)=\tau R_{F,0}(z)\tau^*.
$$

\item[(iii)] If $F\ne0$, then
$$
\sigma_{\mathrm{ess}}(H_{F,\alpha})=\mathbb R.
$$
\end{itemize}
%%%%%%%%%%%%%%%%%%%%%%%%%%%%%%%%%%%%%%
\section{Transmission data and boundary operators}\label{sec:boundary}

\subsection{The free operator and the graph trace}

Fix $d\ge2$ and $F\in\mathbb R$.
Let $\Omega_-\subset\mathbb R^d$ be a bounded Lipschitz domain, put
$$
\Sigma:=\partial\Omega_-,
\qquad
\Omega_+:=\mathbb R^d\setminus\overline{\Omega_-},
$$
and let $\nu$ be the outward unit normal of $\Omega_-$.
The same normal vector $\nu$ is used for the traces from both sides of $\Sigma$.
All inner products and duality pairings are linear in the first argument.
Let
$$
\ell_F=-\Delta-Fx_1
$$
and let $H_{F,0}$ denote the self--adjoint closure of $\ell_F\upharpoonright C_0^\infty(\mathbb R^d)$ in $L^2(\mathbb R^d)$.
We write
$$
D_F:=D(H_{F,0})
$$
and equip $D_F$ with the graph norm
$$
\|u\|_{D_F}:=\|u\|_{L^2(\mathbb R^d)}
+\|H_{F,0}u\|_{L^2(\mathbb R^d)}.
$$
The anti--dual space with respect to the pivot space $L^2(\mathbb R^d)$ is denoted by $D_F'$.

\begin{lem}\label{lem:free-maximal}
The operator $\ell_F\upharpoonright C_0^\infty(\mathbb R^d)$ is essentially self--adjoint, and $C_0^\infty(\mathbb R^d)$ is a graph core for its self--adjoint closure.
Moreover, the closure is the maximal $L^2$ realization of $\ell_F$:
$$
D_F
=
\bigl\{u\in L^2(\mathbb R^d):
\ell_Fu\in L^2(\mathbb R^d)
\text{ in }\mathcal D'(\mathbb R^d)\bigr\},
$$
and $H_{F,0}u=\ell_Fu$ in the distributional sense for $u\in D_F$.
\end{lem}

\begin{proof}
The real potential $V(x)=-Fx_1$ belongs to $L^2_{\mathrm{loc}}(\mathbb R^d)$ and satisfies
$$
V(x)\ge -C_F(1+|x|^2).
$$
The Faris--Lavine theorem therefore implies essential self--adjointness; see \cite[Theorem~X.28]{ReedSimonII}.
Here the initial domain is $C_0^\infty(\mathbb R^d)$.
Let $S_F$ denote the operator on this domain.
A standard distributional characterization of its adjoint gives
$$
D(S_F^*)
=
\bigl\{u\in L^2(\mathbb R^d):
\ell_Fu\in L^2(\mathbb R^d)
\text{ in }\mathcal D'(\mathbb R^d)\bigr\},
\qquad
S_F^*u=\ell_Fu.
$$
Indeed, this follows directly by testing the defining identity for the adjoint against $C_0^\infty(\mathbb R^d)$.
Since $S_F$ is essentially self--adjoint, its closure equals $S_F^*$.
By the definition of the closure, the initial domain $C_0^\infty(\mathbb R^d)$ is a graph core for $H_{F,0}$.
This proves all assertions.
\end{proof}

We recall some standard trace and embedding facts which will be used throughout the paper; see, for example, Adams--Fournier \cite{AdamsFournier2003} and McLean \cite{McLean2000}.
Let $\Omega\subset\mathbb R^d$ be a bounded Lipschitz domain.
Then the trace operator extends to a bounded surjective mapping
$$
\gamma_\Omega:H^1(\Omega)\to H^{1/2}(\partial\Omega),
$$
and it admits a bounded right inverse.

\begin{lem}\label{lem:sobolev-embeddings}
Let $\Sigma\subset\mathbb R^d$ be a compact Lipschitz hypersurface.
Then the embeddings
$$
H^{1/2}(\Sigma)\hookrightarrow L^2(\Sigma),
\qquad
L^2(\Sigma)\hookrightarrow H^{-1/2}(\Sigma)
$$
are continuous.
Moreover, the embedding
$$
H^{1/2}(\Sigma)\hookrightarrow L^2(\Sigma)
$$
is compact.
Consequently,
$$
H^{1/2}(\Sigma)\hookrightarrow H^{-1/2}(\Sigma)
$$
is compact.
\end{lem}

\begin{proof}
The first two embeddings are continuous by the standard Sobolev theory on compact Lipschitz hypersurfaces.
By the Rellich--Kondrachov theorem,
$$
H^{1/2}(\Sigma)\hookrightarrow L^2(\Sigma)
$$
is compact; see \cite{AdamsFournier2003,McLean2000}.
The last statement follows by composition.
\end{proof}

Furthermore, if $u\in H^1(\Omega)$ and $\Delta u\in L^2(\Omega)$ in the distributional sense, then the weak outward normal derivative $\partial_{n_\Omega}u\in H^{-1/2}(\partial\Omega)$ is defined by
$$
\langle\partial_{n_\Omega}u,\phi\rangle
=
\int_\Omega \nabla u\cdot\nabla\overline v\,dx
+
\int_\Omega (\Delta u)\overline v\,dx,
\qquad
\phi=\gamma_\Omega v,
$$
where $v\in H^1(\Omega)$ is any function whose trace is $\phi\in H^{1/2}(\partial\Omega)$.
This definition is independent of the choice of $v$ and yields Green's identity.

We first record the boundedness of the trace map on the domain of the free Stark operator.

\begin{lem}\label{lem:trace-DF}
The trace operator $\tau$ on $\Sigma$ extends to a bounded map
$$
\tau:D_F\to H^{1/2}(\Sigma)
$$
with respect to the graph norm of $H_{F,0}$.
\end{lem}

\begin{proof}
Since $\Sigma$ is compact, there exist bounded open sets
$$
U\Subset U'
$$
such that $\Sigma\subset U$.
Let $u\in D_F$.
Since the action of $H_{F,0}$ agrees with $\ell_F$ in the sense of distributions,
$$
-\Delta u=H_{F,0}u+Fx_1u.
$$
On $U'$ the coordinate function $x_1$ is bounded.
Therefore $-\Delta u\in L^2(U')$.
By local elliptic regularity for the Laplacian, $u\in H^2(U)$, and
$$
\|u\|_{H^2(U)}
\le
C\bigl(
\|u\|_{L^2(U')}+\|\Delta u\|_{L^2(U')}
\bigr).
$$
Using the identity above and the boundedness of $x_1$ on $U'$, we obtain
$$
\|\Delta u\|_{L^2(U')}
\le
C\bigl(
\|H_{F,0}u\|_{L^2(\mathbb R^d)}
+\|u\|_{L^2(\mathbb R^d)}
\bigr).
$$
Hence
$$
\|u\|_{H^2(U)}
\le
C\bigl(
\|u\|_{L^2(\mathbb R^d)}
+\|H_{F,0}u\|_{L^2(\mathbb R^d)}
\bigr).
$$
Choose $\chi\in C_0^\infty(U)$ with $\chi=1$ in a neighborhood of $\Sigma$.
Then $\chi u\in H^2(\mathbb R^d)$, and the global trace theorem on the compact Lipschitz hypersurface gives
$$
\|\tau u\|_{H^{1/2}(\Sigma)}
=
\|\tau(\chi u)\|_{H^{1/2}(\Sigma)}
\le C\|\chi u\|_{H^1(\mathbb R^d)}
\le C\|u\|_{H^2(U)}.
$$
Combining the estimates proves the assertion.
\end{proof}

\begin{lem}\label{lem:tau-star}
By Lemma~\ref{lem:trace-DF}, the adjoint of the graph--norm trace is the bounded map
$$
\tau_D^*:H^{-1/2}(\Sigma)\to D_F',
\qquad
\langle\tau_D^*\varphi,u\rangle_{D_F',D_F}
=
\langle\varphi,\tau u\rangle,
\quad u\in D_F.
$$
The adjoint of the global trace map is the bounded map
$$
\tau_1^*:H^{-1/2}(\Sigma)\to H^{-1}(\mathbb R^d),
\qquad
\langle\tau_1^*\varphi,v\rangle_{H^{-1},H^1}
=
\langle\varphi,\tau v\rangle.
$$
For every $\varphi\in H^{-1/2}(\Sigma)$, the two functionals have the same restriction to $C_0^\infty(\mathbb R^d)$ and hence represent the same distribution, supported on $\Sigma$.
\end{lem}

\begin{proof}
For every compact set $K\subset\mathbb R^d$, the inclusion $C_0^\infty(K)\hookrightarrow D_F$ is continuous: on functions supported in $K$, the graph norm of $H_{F,0}$ is bounded by finitely many of the standard test--function seminorms.
Hence the restriction of $\tau_D^*\varphi$ to $C_0^\infty(\mathbb R^d)$ is a distribution.
Let $\eta\in C_0^\infty(\mathbb R^d)$.
Since $\eta\in D_F\cap H^1(\mathbb R^d)$, the definitions give
$$
\langle\tau_D^*\varphi,\eta\rangle_{D_F',D_F}
=
\langle\varphi,\tau\eta\rangle
=
\langle\tau_1^*\varphi,\eta\rangle_{H^{-1},H^1}.
$$
Thus they induce the same distribution.
If $\eta$ is supported away from $\Sigma$, then $\tau\eta=0$, and the distributional support is contained in $\Sigma$.
\end{proof}

In what follows, $\tau^*$ denotes the graph--dual map $\tau_D^*$ in operator compositions involving $D_F'$, and it denotes the distributional representative $\tau_1^*$ in identities in $\mathcal D'(\mathbb R^d)$.
Lemma~\ref{lem:tau-star} guarantees that these uses are compatible.
No global inclusion $D_F\subset H^1(\mathbb R^d)$ is being asserted.

\begin{lem}\label{lem:tau-dense}
The range of the restricted trace map
$$
\tau\upharpoonright_{D_F}:D_F\to H^{1/2}(\Sigma)
$$
is dense in $H^{1/2}(\Sigma)$.
Consequently, the graph--dual adjoint
$$
\tau_D^*:H^{-1/2}(\Sigma)\to D_F'
$$
is injective.
The global adjoint $\tau_1^*:H^{-1/2}(\Sigma)\to H^{-1}(\mathbb R^d)$ is also injective.
\end{lem}

\begin{proof}
Let $\phi\in H^{1/2}(\Sigma)$.
The trace map on the bounded Lipschitz domain $\Omega_-$ has a bounded right inverse, and $H^1(\Omega_-)$ has a bounded extension operator to $H^1(\mathbb R^d)$; see \cite{AdamsFournier2003,McLean2000}.
Hence the global trace map
$$
\tau:H^1(\mathbb R^d)\to H^{1/2}(\Sigma)
$$
is surjective, and there exists $u\in H^1(\mathbb R^d)$ such that $\tau u=\phi$.
Choose a sequence $u_n\in C_0^\infty(\mathbb R^d)$ satisfying
$$
u_n\to u\quad\text{in }H^1(\mathbb R^d).
$$
Since $C_0^\infty(\mathbb R^d)\subset D_F$, we obtain
$$
\tau u_n\to\tau u=\phi
\quad\text{in }H^{1/2}(\Sigma).
$$
Hence the range is dense.

If $\varphi\in H^{-1/2}(\Sigma)$ satisfies $\tau^*\varphi=0$ in $D_F'$, then
$$
\langle\varphi,\tau u\rangle=0
\qquad\text{for every }u\in D_F.
$$
The density just proved implies $\varphi=0$.
Hence $\tau_D^*$ is injective.
Finally, the global trace is surjective, so its adjoint $\tau_1^*$ is injective as well.
\end{proof}

We now specify the class on which the transmission data are defined.
Choose a ball $B\subset\mathbb R^d$ such that $\overline{\Omega_-}\subset B$ and set
$$
\Omega_+^B:=B\setminus\overline{\Omega_-}.
$$
Already for the classical Laplacian, the generalized Dirichlet and Neumann
traces on the full maximal domain take values in the negative Sobolev spaces
$H^{-1/2}$ and $H^{-3/2}$, respectively; see
\cite[Theorems~8.3.9 and 8.3.10]{BehrndtHassiSnoo2020}.
We therefore do not use the full maximal realization on $\mathbb R^d\setminus\Sigma$.
We instead include piecewise $H^1$ regularity near $\Sigma$ in the definition of the transmission class.
This gives the usual $H^{1/2}$ Dirichlet traces and $H^{-1/2}$ weak normal derivatives without any claim of $H^2$ regularity up to $\Sigma$.

\begin{defn}
We denote by $\mathcal D_\Sigma$ the set of all $u\in L^2(\mathbb R^d)$ with the following properties:
\begin{itemize}
\item[(a)] there exists $g\in L^2(\mathbb R^d)$ such that
$$
\ell_Fu=g
\quad\text{in }\mathcal D'(\mathbb R^d\setminus\Sigma);
$$
\item[(b)] the restrictions satisfy
$$
u^-:=u\upharpoonright_{\Omega_-}\in H^1(\Omega_-),
\qquad
u^+:=u\upharpoonright_{\Omega_+^B}\in H^1(\Omega_+^B).
$$
\end{itemize}
\end{defn}

The function $g$ in (a) is unique as an element of $L^2(\mathbb R^d)$, because two such functions can differ only on $\Sigma$, which has Lebesgue measure zero.
We denote it by
$$
\ell_F^\Sigma u.
$$
For a fixed ball $B$ as above, we use the norm
\begin{eqnarray*}
\|u\|_{\Sigma,B}
&:=&
\|u\|_{L^2(\mathbb R^d)}
+\|\ell_F^\Sigma u\|_{L^2(\mathbb R^d)}
+\|u^-\|_{H^1(\Omega_-)}
+\|u^+\|_{H^1(\Omega_+^B)}.
\end{eqnarray*}

\begin{rem}\label{rem:no-boundary-H2}
The condition
$$
u\in H^2_{\mathrm{loc}}(\mathbb R^d\setminus\Sigma)
$$
does not imply any $H^2$ regularity up to $\Sigma$.
In the definition of $\mathcal D_\Sigma$, the one--sided $H^1$ condition is an
independent assumption.
Interior $H^2$ regularity will be used only on compact sets having positive
distance from $\Sigma$.
Thus no trace on the full maximal domain is used in the construction below.
\end{rem}

\begin{lem}\label{lem:transmission-locality}
Membership in $\mathcal D_\Sigma$ is independent of the particular ball $B$ used in its definition.
Moreover, every $u\in\mathcal D_\Sigma$ satisfies
$$
u\in H^2_{\mathrm{loc}}(\mathbb R^d\setminus\Sigma).
$$
\end{lem}

\begin{proof}
Away from $\Sigma$, condition (a) gives
$$
-\Delta u=\ell_F^\Sigma u+Fx_1u.
$$
On every compact subset of $\mathbb R^d\setminus\Sigma$, the right--hand side belongs to $L^2$, because $x_1$ is locally bounded.
Interior elliptic regularity therefore yields
$$
u\in H^2_{\mathrm{loc}}(\mathbb R^d\setminus\Sigma).
$$

Suppose that condition (b) holds for a ball $B_1$, and let $B_2$ be another ball containing $\overline{\Omega_-}$.
Choose open neighborhoods $V_0,V_1$ of $\Sigma$ such that
$$
\overline{V_0}\subset V_1,
\qquad
\overline{V_1}\subset B_1\cap B_2.
$$
On $V_1\cap\Omega_+$ the required $H^1$ regularity follows from the assumption for $B_1$.
On the other hand,
$$
K:=\overline{\bigl(B_2\setminus\overline{\Omega_-}\bigr)
\setminus V_0}
$$
is a compact subset of $\mathbb R^d\setminus\Sigma$.
The local $H^2$ regularity just proved, together with a finite cover of $K$ and a partition of unity, gives $H^1$ regularity on a neighborhood of $K$.
Combining this with the regularity on $V_1\cap\Omega_+$ yields
$$
u\upharpoonright_{B_2\setminus\overline{\Omega_-}}\in H^1.
$$
Thus condition (b), and hence membership in $\mathcal D_\Sigma$, is independent of $B$.
\end{proof}

\begin{lem}\label{lem:transmission-traces}
Let $u\in\mathcal D_\Sigma$.
Then the Dirichlet traces from the two sides belong to $H^{1/2}(\Sigma)$, and the weak normal derivatives belong to $H^{-1/2}(\Sigma)$.
For every fixed ball $B$ used in the definition of $\mathcal D_\Sigma$, one has
\begin{eqnarray}\label{eq:transmission-trace-estimate}
&&
\|u^-\upharpoonright_\Sigma\|_{H^{1/2}(\Sigma)}
+\|u^+\upharpoonright_\Sigma\|_{H^{1/2}(\Sigma)}
\nonumber\\
&&\quad
+\|(\partial_\nu u)^-\|_{H^{-1/2}(\Sigma)}
+\|(\partial_\nu u)^+\|_{H^{-1/2}(\Sigma)}
\le C_B\|u\|_{\Sigma,B}.
\end{eqnarray}
The exterior weak normal derivative and all transmission data are independent of the auxiliary ball.
\end{lem}

\begin{proof}
The trace theorem on the bounded Lipschitz domains $\Omega_-$ and $\Omega_+^B$ gives
$$
u^\pm\upharpoonright_\Sigma\in H^{1/2}(\Sigma)
$$
and the required estimates for the Dirichlet traces.
On either side of $\Sigma$, the equation
$$
-\Delta u^\pm=\ell_F^\Sigma u+Fx_1u
$$
holds in the sense of distributions.
Since $x_1$ is bounded on $B$, the right--hand side belongs to $L^2$ there.
For a bounded Lipschitz domain $D$, the weak normal derivative satisfies
\begin{eqnarray*}
\|\partial_{n_D}v\|_{H^{-1/2}(\partial D)}
&\le&
C_D\bigl(
\|v\|_{H^1(D)}+\|\Delta v\|_{L^2(D)}
\bigr)
\end{eqnarray*}
whenever $v\in H^1(D)$ and $\Delta v\in L^2(D)$.
Application of this estimate on $\Omega_-$ and $\Omega_+^B$ proves \eqref{eq:transmission-trace-estimate}.
For clarity, the exterior normal derivative is defined as follows.
For $\phi\in H^{1/2}(\Sigma)$, choose
$V\in H^1(\Omega_+^B)$ whose trace is $\phi$ on $\Sigma$ and zero on
$\partial B$.
Then
\begin{eqnarray*}
\langle(\partial_\nu u)^+,\phi\rangle
&:=&
-\int_{\Omega_+^B}\nabla u^+\cdot\nabla\overline V\,dx
-\int_{\Omega_+^B}(\Delta u^+)\overline V\,dx.
\end{eqnarray*}
This is the negative of the weak outward normal derivative of
$\Omega_+^B$ on $\Sigma$, because the outward normal of the exterior piece is
$-\nu$ there.
The definition is independent of the choice of $V$.
If two balls are used, one may choose an extension supported in a common
neighborhood of $\Sigma$; the two formulas then agree.
The resulting functional on $H^{1/2}(\Sigma)$ is therefore independent of the
ball.
The same is clear for the Dirichlet traces.
\end{proof}

Thus the traces from the two sides
$$
u^\pm\upharpoonright_\Sigma\in H^{1/2}(\Sigma)
$$
and the weak normal derivatives are well defined.
We put
$$
[u]:=u^+\upharpoonright_\Sigma-u^-\upharpoonright_\Sigma,
\qquad
[\partial_\nu u]:=(\partial_\nu u)^+-(\partial_\nu u)^-.
$$
If $[u]=0$, their common trace is denoted by
$$
\gamma u:=u^+\upharpoonright_\Sigma
=u^-\upharpoonright_\Sigma\in H^{1/2}(\Sigma).
$$

\begin{lem}\label{lem:free-in-transmission}
If $u\in D_F$, then $u\in\mathcal D_\Sigma$ and
$$
\ell_F^\Sigma u=H_{F,0}u,
\qquad
[u]=0,
\qquad
[\partial_\nu u]=0.
$$
In particular, the common trace satisfies $\gamma u=\tau u$.
\end{lem}

\begin{proof}
Condition (a) in the definition of $\mathcal D_\Sigma$ follows from Lemma~\ref{lem:free-maximal}.
Choose a ball $B$ containing $\overline{\Omega_-}$ and a larger ball $B'$ such that $\overline B\subset B'$.
On $B'$ one has
$$
-\Delta u=H_{F,0}u+Fx_1u\in L^2(B').
$$
Interior elliptic regularity therefore gives $u\in H^2(B)$.
Hence the restrictions of $u$ to $\Omega_-$ and $\Omega_+^B$ belong to $H^1$, so condition (b) holds.
Since $u$ is a single $H^2$ function across $\Sigma$ in a neighborhood of the interface, its one--sided traces and its weak normal derivatives, computed with the same fixed normal $\nu$, coincide.
Thus both jumps vanish, and the common trace is precisely the graph--norm trace $\tau u$ from Lemma~\ref{lem:trace-DF}.
\end{proof}

\begin{lem}\label{lem:cutoff-DF}
Let $u\in\mathcal D_\Sigma$, and let $\chi\in C^\infty(\mathbb R^d)$ be bounded with bounded derivatives, such that $\chi=0$ in a neighborhood of $\Sigma$ and the derivatives of $\chi$ have compact support.
Then $\chi u\in D_F$.
\end{lem}

\begin{proof}
Since $u\in H^2_{\mathrm{loc}}(\mathbb R^d\setminus\Sigma)$ and the support of $\nabla\chi$ is compact and disjoint from $\Sigma$, the product rule gives
$$
\ell_F(\chi u)
=
\chi\ell_F^\Sigma u
-2\nabla\chi\cdot\nabla u
-(\Delta\chi)u
$$
in the sense of distributions.
Every term on the right belongs to $L^2(\mathbb R^d)$: the first because $\chi$ is bounded and $\ell_F^\Sigma u\in L^2$, and the remaining terms because their supports are compact and separated from $\Sigma$, where $u\in H^2_{\mathrm{loc}}$.
Lemma~\ref{lem:free-maximal} now implies $\chi u\in D_F$.
\end{proof}

\begin{lem}\label{lem:distributional-jump}
Let $u\in\mathcal D_\Sigma$ and suppose that $[u]=0$.
Then
\begin{equation}\label{eq:distributional-jump}
\ell_Fu
=
\ell_F^\Sigma u-\tau_1^*[\partial_\nu u]
\quad\text{in }\mathcal D'(\mathbb R^d).
\end{equation}
Equivalently, for every $z\in\mathbb C$,
$$
(\ell_F-z)u
=
(\ell_F^\Sigma u-zu)-\tau_1^*[\partial_\nu u].
$$
\end{lem}

\begin{proof}
Let $\eta\in C_0^\infty(\mathbb R^d)$, and choose a ball $B$ containing $\overline{\Omega_-}$ and $\operatorname{supp}\eta$.
Put $D_+=B\setminus\overline{\Omega_-}$.
The test function vanishes near $\partial B$.
Integration by parts on $\Omega_-$ gives
$$
\langle\ell_Fu,\eta\rangle_{\Omega_-}
=(\ell_F^\Sigma u,\eta)_{L^2(\Omega_-)}
+\langle(\partial_\nu u)^-,\tau\eta\rangle
-\int_\Sigma (u^-)\,\overline{\partial_\nu\eta}\,d\sigma.
$$
The outward normal of $D_+$ equals $-\nu$ on $\Sigma$, and hence
$$
\langle\ell_Fu,\eta\rangle_{D_+}
=(\ell_F^\Sigma u,\eta)_{L^2(D_+)}
-\langle(\partial_\nu u)^+,\tau\eta\rangle
+\int_\Sigma (u^+)\,\overline{\partial_\nu\eta}\,d\sigma.
$$
Adding the two formulas yields
$$
\langle\ell_Fu,\eta\rangle
=
(\ell_F^\Sigma u,\eta)_{L^2}
-\langle[\partial_\nu u],\tau\eta\rangle
+\int_\Sigma [u]\,\overline{\partial_\nu\eta}\,d\sigma.
$$
The last integral vanishes because $[u]=0$.
The multiplication potential $-Fx_1$ contributes only to the volume terms, so the displayed identity is exactly \eqref{eq:distributional-jump}.
In particular, the minus sign in front of the normal jump is fixed by the chosen convention that $\nu$ points from $\Omega_-$ to $\Omega_+$.
\end{proof}

\begin{lem}[Green identity]\label{lem:green-transmission}
Let $u,v\in\mathcal D_\Sigma$ and suppose that $[u]=[v]=0$.
Then
\begin{equation}\label{eq:green-transmission}
(\ell_F^\Sigma u,v)_{L^2}
-(u,\ell_F^\Sigma v)_{L^2}
=
\langle[\partial_\nu u],\gamma v\rangle
-\overline{\langle[\partial_\nu v],\gamma u\rangle}.
\end{equation}
\end{lem}

\begin{proof}
Choose a ball $B$ as above and $\chi\in C_0^\infty(B)$ such that $\chi=1$ in a neighborhood of $\Sigma$.
Write
$$
u=u_c+u_\infty,\qquad v=v_c+v_\infty,
$$
where
$$
u_c=\chi u,\quad u_\infty=(1-\chi)u,
\qquad
v_c=\chi v,\quad v_\infty=(1-\chi)v.
$$
By Lemma~\ref{lem:cutoff-DF}, $u_\infty,v_\infty\in D_F$, and their traces on $\Sigma$ vanish.
The compactly supported functions $u_c,v_c$ have the same traces and normal jumps at $\Sigma$ as $u,v$.

Apply the second weak Green identity first on $\Omega_-$ and then on $B\setminus\overline{\Omega_-}$.
The compactly supported pieces vanish near $\partial B$.
On $\Sigma$, the outward normal of the exterior piece is $-\nu$.
Since $[u_c]=[v_c]=0$, the two boundary contributions combine to give
\begin{equation}\label{eq:green-compact}
(\ell_F^\Sigma u_c,v_c)
-(u_c,\ell_F^\Sigma v_c)
=
\langle[\partial_\nu u],\gamma v\rangle
-\overline{\langle[\partial_\nu v],\gamma u\rangle}.
\end{equation}
Notice that the real potential $-Fx_1$ cancels from the skew part.

We next verify the cancellation of all cross terms.
Since $C_0^\infty(\mathbb R^d)$ is a graph core for $H_{F,0}$, choose $\eta_n\in C_0^\infty(\mathbb R^d)$ such that $\eta_n\to v_\infty$ in $D_F$.
By Lemma~\ref{lem:distributional-jump},
$$
(\ell_F^\Sigma u_c,\eta_n)
-\langle[\partial_\nu u],\tau\eta_n\rangle
=(u_c,H_{F,0}\eta_n).
$$
The graph--norm continuity of the trace and $\tau v_\infty=0$ imply $\tau\eta_n\to0$ in $H^{1/2}(\Sigma)$.
Passing to the limit gives
\begin{equation}\label{eq:cross-one}
(\ell_F^\Sigma u_c,v_\infty)
=(u_c,H_{F,0}v_\infty).
\end{equation}
Similarly, approximate $u_\infty$ in the graph norm by $\xi_n\in C_0^\infty(\mathbb R^d)$.
Applying the distributional jump formula to $v_c$, then taking complex conjugates and passing to the limit, yields
\begin{equation}\label{eq:cross-two}
(H_{F,0}u_\infty,v_c)
=(u_\infty,\ell_F^\Sigma v_c).
\end{equation}
Finally, self--adjointness of $H_{F,0}$ gives
\begin{equation}\label{eq:cross-three}
(H_{F,0}u_\infty,v_\infty)
=(u_\infty,H_{F,0}v_\infty).
\end{equation}
Since
$$
\ell_F^\Sigma u=\ell_F^\Sigma u_c+H_{F,0}u_\infty,
\qquad
\ell_F^\Sigma v=\ell_F^\Sigma v_c+H_{F,0}v_\infty,
$$
expansion of the left-hand side of \eqref{eq:green-transmission}, followed by the three identities \eqref{eq:cross-one}, \eqref{eq:cross-two}, and \eqref{eq:cross-three}, leaves only the compact part \eqref{eq:green-compact}.
This proves the identity.
\end{proof}

%%%%%%%%%%%%%%%%%%%%%%%%%%%
\subsection{Definition of $H_{F,\alpha}$ (operator realization)}

Let $\alpha\in L^\infty(\Sigma;\mathbb R)$.
We regard $\alpha$ as the multiplication operator
$$
\alpha:H^{1/2}(\Sigma)\to H^{-1/2}(\Sigma),
\qquad \phi\mapsto\alpha\phi,
$$
where $\alpha\phi\in L^2(\Sigma)$ is identified with an element of $H^{-1/2}(\Sigma)$ by the canonical embedding $L^2(\Sigma)\hookrightarrow H^{-1/2}(\Sigma)$.

We define
\begin{equation}\label{eq:domain-HFa}
D(H_{F,\alpha})
=
\Bigl\{u\in\mathcal D_\Sigma:
[u]=0\text{ in }H^{1/2}(\Sigma),\quad
[\partial_\nu u]=\alpha\gamma u
\text{ in }H^{-1/2}(\Sigma)
\Bigr\},
\end{equation}
and
$$
H_{F,\alpha}u:=\ell_F^\Sigma u,
\qquad u\in D(H_{F,\alpha}).
$$
Thus the operator acts by the Stark differential expression away from the interface and the singular interaction is encoded by the transmission conditions.

We next verify that $H_{F,\alpha}$ is symmetric.

\begin{lem}\label{lem:symmetric-HFa}
The operator $H_{F,\alpha}$ is densely defined in $L^2(\mathbb R^d)$ and is symmetric.
\end{lem}

\begin{proof}
The inclusion
$$
C_0^\infty(\mathbb R^d\setminus\Sigma)
\subset D(H_{F,\alpha})
$$
shows that the domain is dense.
Indeed, since $\Sigma$ has Lebesgue measure zero, $C_0^\infty(\mathbb R^d\setminus\Sigma)$ is dense in $L^2(\mathbb R^d)$.
Let $u,v\in D(H_{F,\alpha})$.
By Lemma~\ref{lem:green-transmission},
$$
(H_{F,\alpha}u,v)-(u,H_{F,\alpha}v)
=
\langle\alpha\gamma u,\gamma v\rangle
-\overline{\langle\alpha\gamma v,\gamma u\rangle}.
$$
Because $\alpha$ is real valued, multiplication by $\alpha$ is symmetric on $L^2(\Sigma)$, and the right--hand side vanishes.
Hence $H_{F,\alpha}$ is symmetric.
\end{proof}

\begin{rem}
When $F=0$, the domain condition \eqref{eq:domain-HFa} is the standard transmission condition for Schr\"odinger operators with $\delta$ interactions supported on $\Sigma$.
For $F\ne0$ we use the same interface condition, while the background resolvent and the operator domain are those of the Stark Hamiltonian.
The self--adjointness of this realization will be obtained below from the resolvent formula.
\end{rem}

%%%%%%%%%%%%%%%%%%%%%%%%%
\subsection{Boundary integral operators and resolvent representation}

Let
$$
R_{F,0}(z)=(H_{F,0}-z)^{-1},
\qquad z\in\mathbb C\setminus\mathbb R.
$$
We use the following standard extension to the graph anti--dual.

\begin{lem}\label{lem:dual-resolvent}
For every $z\in\mathbb C\setminus\mathbb R$, the free resolvent has a unique bounded extension
$$
R_{F,0}(z):D_F'\to L^2(\mathbb R^d)
$$
defined by
\begin{equation}\label{eq:dual-resolvent}
(R_{F,0}(z)g,f)_{L^2}
=
\langle g,R_{F,0}(\overline z)f\rangle_{D_F',D_F},
\qquad f\in L^2(\mathbb R^d).
\end{equation}
If $w=R_{F,0}(z)g$, then
$$
(H_{F,0}-z)w=g
$$
in $D_F'$, where the transposed action is defined by
$$
\bigl\langle(H_{F,0}-z)w,u\bigr\rangle_{D_F',D_F}
:=
(w,(H_{F,0}-\overline z)u)_{L^2},
\qquad u\in D_F.
$$
If $g$ has a distributional representative, the same equality holds in $\mathcal D'(\mathbb R^d)$.
\end{lem}

\begin{proof}
The map $R_{F,0}(\overline z):L^2\to D_F$ is bounded.
Hence the right--hand side of \eqref{eq:dual-resolvent} is a bounded anti--linear functional of $f\in L^2$, and the Riesz representation theorem defines a unique $w\in L^2$ with the stated bound.
For $u\in D_F$, take $f=(H_{F,0}-\overline z)u$ in \eqref{eq:dual-resolvent}.
Since $R_{F,0}(\overline z)(H_{F,0}-\overline z)u=u$, one obtains
$$
(w,(H_{F,0}-\overline z)u)_{L^2}
=\langle g,u\rangle_{D_F',D_F},
$$
which is precisely the transposed identity $(H_{F,0}-z)w=g$.
Agreement on test functions gives the distributional statement.
\end{proof}

Since $\tau^*:H^{-1/2}(\Sigma)\to D_F'$ is bounded, the expression
$$
G_F(z)\varphi:=R_{F,0}(z)\tau^*\varphi
$$
is well defined as an element of $L^2(\mathbb R^d)$ for every $\varphi\in H^{-1/2}(\Sigma)$.

The classical single--layer construction for the Laplacian on a Lipschitz
boundary will be used only for the local comparison problem below; see
\cite{Costabel1988,McLean2000,Verchota1984}.
The mapping properties of the Stark layer potential are then derived from the
free Stark resolvent and local elliptic regularity rather than from a direct
application of translation invariant layer potential theory.

\begin{lem}\label{lem:single--layer}
Let $z\in\mathbb C\setminus\mathbb R$ and let $\varphi\in H^{-1/2}(\Sigma)$.
Define
$$
w:=G_F(z)\varphi=R_{F,0}(z)\tau^*\varphi.
$$
Then the following statements hold.
\begin{itemize}
\item[(i)]
$(\ell_F-z)w=\tau_1^*\varphi$ in $\mathcal D'(\mathbb R^d)$, and $(\ell_F-z)w=0$ in $\mathcal D'(\mathbb R^d\setminus\Sigma)$.

\item[(ii)]
$w\in\mathcal D_\Sigma$ and $w\in H^2_{\mathrm{loc}}(\mathbb R^d\setminus\Sigma)$.

\item[(iii)] The traces satisfy
$$
[w]=0,\qquad [\partial_\nu w]=-\varphi,
$$
and the common trace $\gamma w$ belongs to $H^{1/2}(\Sigma)$.

\item[(iv)] The map
$$
H^{-1/2}(\Sigma)\ni\varphi
\mapsto\gamma w\in H^{1/2}(\Sigma)
$$
is linear and bounded.
We denote this operator by
$$
M_F(z)\varphi:=\gamma w.
$$
\end{itemize}
\end{lem}

\begin{proof}
By Lemma~\ref{lem:dual-resolvent} and Lemma~\ref{lem:tau-star},
$$
(\ell_F-z)w=\tau_1^*\varphi
$$
holds in $\mathcal D'(\mathbb R^d)$.
Since $\operatorname{supp}(\tau_1^*\varphi)\subset\Sigma$, the equation is homogeneous on $\mathbb R^d\setminus\Sigma$.
For every compact $K\Subset\mathbb R^d\setminus\Sigma$, the function $x_1$ is bounded on $K$ and
$$
-\Delta w=(z+Fx_1)w\in L^2(K).
$$
Interior elliptic regularity gives $w\in H^2_{\mathrm{loc}}(\mathbb R^d\setminus\Sigma)$.

To obtain regularity up to and across the interface, fix $\lambda>0$ and let $s\in H^1(\mathbb R^d)$ be the unique weak solution of
\begin{equation}\label{eq:comparison-layer}
(-\Delta+\lambda)s=\tau_1^*\varphi
\quad\text{in }H^{-1}(\mathbb R^d).
\end{equation}
Existence, uniqueness, and the estimate
$$
\|s\|_{H^1(\mathbb R^d)}
\le C_\lambda\|\varphi\|_{H^{-1/2}(\Sigma)}
$$
follow from the Lax--Milgram theorem and the boundedness of the global trace.
The source in \eqref{eq:comparison-layer} is supported on $\Sigma$, so $s\in H^2_{\mathrm{loc}}(\mathbb R^d\setminus\Sigma)$.
Because $s$ is a single global $H^1$ function, its two traces agree, and
$$
\|\tau s\|_{H^{1/2}(\Sigma)}
\le C\|\varphi\|_{H^{-1/2}(\Sigma)}.
$$
This is the usual weak single--layer construction.
Classical treatments on Lipschitz boundaries may be found in \cite{Costabel1988,McLean2000,Verchota1984}.

Set $v=w-s$.
Subtracting \eqref{eq:comparison-layer} from the equation for $w$ gives
\begin{equation}\label{eq:v-equation}
(-\Delta-Fx_1-z)v
=(\lambda+z+Fx_1)s
\quad\text{in }\mathcal D'(\mathbb R^d).
\end{equation}
The right--hand side belongs to $L^2_{\mathrm{loc}}(\mathbb R^d)$.
Since $v\in L^2_{\mathrm{loc}}(\mathbb R^d)$ and $x_1$ is locally bounded,
\begin{eqnarray*}
-\Delta v
&=&
(Fx_1+z)v+(\lambda+z+Fx_1)s
\in L^2_{\mathrm{loc}}(\mathbb R^d).
\end{eqnarray*}
Local elliptic regularity therefore yields
$v\in H^2_{\mathrm{loc}}(\mathbb R^d)$, now across $\Sigma$ as well.
Thus $w=s+v$ belongs to $H^1$ in a neighborhood of $\Sigma$, and its two traces agree.
Together with the already established $H^2_{\mathrm{loc}}$ regularity away from $\Sigma$, a finite--cover argument on any ball containing $\overline{\Omega_-}$ shows that the two restrictions required in condition (b) of the definition of $\mathcal D_\Sigma$ belong to $H^1$.
Condition (a) holds with $g=zw\in L^2$.
Consequently $w\in\mathcal D_\Sigma$, and its regular action is $\ell_F^\Sigma w=zw$.

Lemma~\ref{lem:distributional-jump} now gives
$$
(\ell_F-z)w=-\tau_1^*[\partial_\nu w].
$$
Comparison with $(\ell_F-z)w=\tau_1^*\varphi$ yields
$$
\tau_1^*\bigl(\varphi+[\partial_\nu w]\bigr)=0.
$$
The injectivity of $\tau_1^*$, recorded in Lemma~\ref{lem:tau-dense}, implies
$$
[\partial_\nu w]=-\varphi.
$$
This proves assertions (i), (ii), and (iii), including the sign.

For completeness, we establish the trace estimate.
Choose bounded open sets $U\Subset U'$ with $\Sigma\subset U$.
A local elliptic estimate applied to \eqref{eq:v-equation} gives
$$
\|v\|_{H^2(U)}
\le C\left(
\|v\|_{L^2(U')}
+\|(\lambda+z+Fx_1)s\|_{L^2(U')}
\right).
$$
By Lemma~\ref{lem:dual-resolvent},
$$
\|w\|_{L^2}
\le C\|\tau_D^*\varphi\|_{D_F'}
\le C\|\varphi\|_{H^{-1/2}(\Sigma)},
$$
and the same type of bound holds for $s$ in $H^1$.
Hence
$$
\|v\|_{H^2(U)}
\le C\|\varphi\|_{H^{-1/2}(\Sigma)}.
$$
The trace theorem and the estimate for $s$ imply
$$
\|\gamma w\|_{H^{1/2}(\Sigma)}
\le C\|\varphi\|_{H^{-1/2}(\Sigma)}.
$$
This proves (iv).
\end{proof}

\begin{rem}[Sign check]\label{rem:sign-check}
The sign in Lemma~\ref{lem:single--layer} can be checked in one dimension.
If a continuous, piecewise $H^2$ function has derivative jump $[u']=u'(0+)-u'(0-)$, then
$$
-u''=(-u'')_{\mathrm{reg}}-[u']\,\delta_0
$$
in distributions.
Thus a positive source $\varphi\delta_0$ produces the jump $[u']=-\varphi$.
This is exactly the convention used above and is the reason for both the plus sign in $I+\alpha M_F(z)$ and the minus sign in the resolvent correction.
\end{rem}

We define the boundary operator
$$
M_F(z):H^{-1/2}(\Sigma)\to H^{1/2}(\Sigma),
\qquad
M_F(z)\varphi
:=\gamma\bigl(R_{F,0}(z)\tau^*\varphi\bigr).
$$
The familiar notation $M_F(z)=\tau R_{F,0}(z)\tau^*$ is used only as a formal shorthand.
The trace on the left is the common transmission trace $\gamma$ of the single--layer potential; it is not an application of the graph--domain trace $\tau:D_F\to H^{1/2}(\Sigma)$ to a vector that need not belong to $D_F$.

For later use, we also introduce the boundary operator
\begin{equation}\label{eq:B-alpha}
B_\alpha(z):=I+\alpha M_F(z)
\quad\text{on }H^{-1/2}(\Sigma).
\end{equation}
The order of the factors and the plus sign in this definition are fixed by the jump convention in Lemma~\ref{lem:single--layer}.

\begin{rem}\label{rem:factorization}
The Birman--Schwinger reduction reflects a natural factorization of the hypersurface interaction.
At the level of forms one may write
$$
\alpha\delta_\Sigma=\tau_1^*\alpha\tau.
$$
Indeed, for $u,v\in H^1(\mathbb R^d)$,
$$
\langle\tau_1^*\alpha\tau u,v\rangle
=
\langle\alpha\tau u,\tau v\rangle
=
\int_\Sigma\alpha(\tau u)\overline{\tau v}\,d\sigma.
$$
By Lemma~\ref{lem:distributional-jump}, the transmission condition $[\partial_\nu u]=\alpha\gamma u$ exactly cancels the surface distribution in the formal expression $\ell_Fu+\tau_1^*\alpha\gamma u$.
Accordingly, for the single--layer ansatz at $z\in\mathbb C\setminus\mathbb R$, the boundary equation is
$$
B_\alpha(z)\varphi=(I+\alpha M_F(z))\varphi=0.
$$
The same sign convention produces the minus sign in the resolvent formula below.
\end{rem}

We now state the resolvent formula for $H_{F,\alpha}$.

\begin{thm}\label{thm:resolvent-new}
Let $\alpha\in L^\infty(\Sigma;\mathbb R)$ and $z\in\mathbb C\setminus\mathbb R$.
Then the operator $B_\alpha(z)$ in \eqref{eq:B-alpha} is boundedly invertible on $H^{-1/2}(\Sigma)$, and
\begin{equation}\label{eq:resolvent-new}
(H_{F,\alpha}-z)^{-1}
=
R_{F,0}(z)
-
R_{F,0}(z)\tau^*(I+\alpha M_F(z))^{-1}
\alpha\tau R_{F,0}(z).
\end{equation}
\end{thm}

\begin{proof}
We first show that
$$
B_\alpha(z)=I+\alpha M_F(z):H^{-1/2}(\Sigma)
\to H^{-1/2}(\Sigma)
$$
is invertible.
The operator
$$
M_F(z):H^{-1/2}(\Sigma)\to H^{1/2}(\Sigma)
$$
is bounded.
By Lemma~\ref{lem:sobolev-embeddings}, multiplication by $\alpha$ defines a compact map
$$
\alpha:H^{1/2}(\Sigma)\to H^{-1/2}(\Sigma).
$$
Therefore $\alpha M_F(z)$ is compact on $H^{-1/2}(\Sigma)$ and $B_\alpha(z)$ is Fredholm of index zero.

It remains to prove injectivity.
Assume that
$$
B_\alpha(z)\varphi=0
$$
for some $\varphi\in H^{-1/2}(\Sigma)$, and define
$$
w:=R_{F,0}(z)\tau^*\varphi.
$$
By Lemma~\ref{lem:single--layer},
$$
(\ell_F-z)w=0
\quad\text{in }\mathbb R^d\setminus\Sigma,
\qquad
[w]=0,
\qquad
[\partial_\nu w]=-\varphi,
\qquad
\gamma w=M_F(z)\varphi.
$$
The boundary equation implies
$$
-\varphi=\alpha M_F(z)\varphi=\alpha\gamma w.
$$
Hence $w\in D(H_{F,\alpha})$ and
$$
(H_{F,\alpha}-z)w=0.
$$
By Lemma~\ref{lem:symmetric-HFa}, $(H_{F,\alpha}w,w)$ is real.
Taking imaginary parts in
$$
0=((H_{F,\alpha}-z)w,w)
$$
gives
$$
-\operatorname{Im}z\,\|w\|^2=0.
$$
Since $\operatorname{Im}z\ne0$, it follows that $w=0$.
On the other hand,
$$
(\ell_F-z)w=\tau_1^*\varphi
$$
in the sense of distributions.
Thus $\tau_1^*\varphi=0$, and Lemma~\ref{lem:tau-dense} yields $\varphi=0$.
Since $B_\alpha(z)$ is Fredholm of index zero, injectivity implies surjectivity.
Its inverse is bounded by the bounded inverse theorem.

Next let $f\in L^2(\mathbb R^d)$ and put
$$
u_0:=R_{F,0}(z)f\in D_F.
$$
By Lemma~\ref{lem:free-in-transmission}, $u_0\in\mathcal D_\Sigma$, its two jumps vanish, and $\gamma u_0=\tau u_0\in H^{1/2}(\Sigma)$.
Define
$$
\psi:=B_\alpha(z)^{-1}\alpha\tau u_0
\in H^{-1/2}(\Sigma)
$$
and set
$$
u:=u_0-G_F(z)\psi.
$$
The class $\mathcal D_\Sigma$ is linear, and both terms in this difference belong to it by Lemma~\ref{lem:free-in-transmission} and Lemma~\ref{lem:single--layer}.
Moreover, Lemma~\ref{lem:single--layer} gives
$$
[u]=0,
\qquad
[\partial_\nu u]=\psi,
\qquad
\gamma u=\tau u_0-M_F(z)\psi.
$$
The equation $B_\alpha(z)\psi=\alpha\tau u_0$ is equivalent to
$$
\psi
=\alpha\bigl(\tau u_0-M_F(z)\psi\bigr)
=\alpha\gamma u.
$$
Moreover,
$$
\ell_F^\Sigma u
=
\ell_Fu_0-zG_F(z)\psi
=
f+zu_0-zG_F(z)\psi
=
f+zu
\in L^2(\mathbb R^d).
$$
Thus $u\in D(H_{F,\alpha})$ and
$$
(H_{F,\alpha}-z)u=f.
$$
Therefore $H_{F,\alpha}-z$ is surjective.
This construction also displays the signs in the claimed resolvent formula without introducing a negative boundary density.

To prove injectivity, let $u\in D(H_{F,\alpha})$ satisfy $(H_{F,\alpha}-z)u=0$.
Since $H_{F,\alpha}$ is symmetric,
$$
0=((H_{F,\alpha}-z)u,u)
=(H_{F,\alpha}u,u)-z\|u\|^2.
$$
Taking imaginary parts gives $-\operatorname{Im}z\,\|u\|^2=0$, and hence $u=0$.
Thus $H_{F,\alpha}-z$ is bijective.

For the unique solution constructed above,
\begin{eqnarray*}
u
&=&R_{F,0}(z)f
-R_{F,0}(z)\tau^*(I+\alpha M_F(z))^{-1}
\alpha\tau R_{F,0}(z)f.
\end{eqnarray*}
Every factor on the right--hand side of \eqref{eq:resolvent-new} is bounded between the indicated spaces.
Hence the displayed solution operator is bounded on $L^2(\mathbb R^d)$.
The surjectivity and injectivity proved above show that it is the two--sided inverse of $H_{F,\alpha}-z$.
This proves \eqref{eq:resolvent-new}.
\end{proof}

As an immediate consequence of the resolvent formula, we obtain the self--adjointness of $H_{F,\alpha}$.

\begin{cor}\label{cor:selfadjoint-HFa}
The operator defined in \eqref{eq:domain-HFa} is self--adjoint in $L^2(\mathbb R^d)$.
\end{cor}

\begin{proof}
By Lemma~\ref{lem:symmetric-HFa}, $H_{F,\alpha}$ is densely defined and symmetric.
Theorem~\ref{thm:resolvent-new}, applied to $z=i$ and $z=-i$, gives
$$
\operatorname{Ran}(H_{F,\alpha}-i)
=
\operatorname{Ran}(H_{F,\alpha}+i)
=
L^2(\mathbb R^d).
$$
The standard range criterion for densely defined symmetric operators therefore
implies that $H_{F,\alpha}$ is self--adjoint.
\end{proof}

%%%%%%%%%%%%%%%%%%%%%%%%%
\section{A spectral application: preservation of the essential spectrum}

The main objective of the present work is to establish a boundary resolvent framework for Stark Hamiltonians with $\delta$--interactions supported on a compact hypersurface.
The results obtained in Section~\ref{sec:boundary} express the resolvent of the interacting operator in terms of the free Stark resolvent and a boundary operator acting on $\Sigma$.
As a first spectral application of this representation, we analyze the essential spectrum of the operator in the presence of the electric field.

For comparison, in the zero--field case one has a much sharper spectral and scattering picture in special geometries; see, in particular, Ikebe and Shimada~\cite{IkebeShimada1991} for penetrable--wall Hamiltonians.
We focus here on the Stark case and on the general compact Lipschitz hypersurface setting, where the argument below yields compactness of the resolvent difference, which is sufficient for the preservation of the essential spectrum.
The compactness argument itself is valid for every $F\in\mathbb R$.
For the boundary $\Sigma$ of a bounded Lipschitz domain, the boundary resolvent formula obtained in Section~\ref{sec:boundary} shows that, for every $z\in\mathbb C\setminus\mathbb R$, the resolvent difference
$$
(H_{F,\alpha}-z)^{-1}-(H_{F,0}-z)^{-1}
$$
is compact on $L^2(\mathbb R^d)$.
For $F\ne0$, this preserves the free essential spectrum $\mathbb R$.

We first isolate the compactness of the resolvent difference.
The spectral consequence is then immediate from Weyl's theorem.

\begin{prop}\label{prop:compact-resolvent-difference}
Let $\Sigma\subset\mathbb R^d$ be the boundary of a bounded Lipschitz domain, let $\alpha\in L^\infty(\Sigma;\mathbb R)$, let $d\ge2$, and let $F\in\mathbb R$.
Then for every $z\in\mathbb C\setminus\mathbb R$ the resolvent difference
$$
(H_{F,\alpha}-z)^{-1}-(H_{F,0}-z)^{-1}
$$
is compact on $L^2(\mathbb R^d)$.
\end{prop}

\begin{proof}
Set
$$
H:=H_{F,\alpha},\qquad H_0:=H_{F,0}.
$$
Fix $z\in\mathbb C\setminus\mathbb R$.
By the resolvent formula in Theorem~\ref{thm:resolvent-new},
\begin{equation}\label{eq:resdiff}
(H-z)^{-1}-(H_0-z)^{-1}
=
-R_{F,0}(z)\tau^*B_\alpha(z)^{-1}
\alpha\tau R_{F,0}(z).
\end{equation}

Now
$$
R_{F,0}(z):L^2(\mathbb R^d)\to D_F
$$
is bounded.
Since the trace map
$$
\tau:D_F\to H^{1/2}(\Sigma)
$$
is bounded by Lemma~\ref{lem:trace-DF}, it follows that
$$
\tau R_{F,0}(z):L^2(\mathbb R^d)\to H^{1/2}(\Sigma)
$$
is bounded.
By Lemma~\ref{lem:sobolev-embeddings}, the embedding
$$
H^{1/2}(\Sigma)\hookrightarrow L^2(\Sigma)
$$
is compact.
Since multiplication by $\alpha\in L^\infty(\Sigma;\mathbb R)$ defines a bounded operator on $L^2(\Sigma)$ and the canonical embedding $L^2(\Sigma)\hookrightarrow H^{-1/2}(\Sigma)$ is continuous, the map
$$
\alpha:H^{1/2}(\Sigma)\to H^{-1/2}(\Sigma)
$$
is compact.
Finally,
$$
B_\alpha(z)^{-1}:H^{-1/2}(\Sigma)\to H^{-1/2}(\Sigma)
$$
and
$$
R_{F,0}(z)\tau^*:H^{-1/2}(\Sigma)\to L^2(\mathbb R^d)
$$
are bounded.
Hence the operator on the right--hand side of \eqref{eq:resdiff} is compact.
\end{proof}

\begin{rem}
The above argument determines the essential spectrum of $H_{F,\alpha}$, but it does not determine the spectral type.
In particular, we do not address the absence of singular continuous spectrum or embedded eigenvalues.
In the zero--field case, substantially more detailed spectral and scattering results are known in special geometries; see, in particular, Ikebe and Shimada~\cite{IkebeShimada1991} for penetrable--wall Hamiltonians.
In the present paper, by contrast, we work in the general framework of compact Lipschitz hypersurfaces and focus on the Stark case $F\ne0$.
For this purpose, compactness of the resolvent difference is sufficient to obtain the preservation of the essential spectrum.
A more detailed analysis of the spectral type in the Stark case would require additional tools from scattering theory or commutator methods.
Moreover, the hypersurface--supported $\delta$--interaction makes the commutator analysis for a Mourre estimate more delicate, and we do not pursue this question here.
\end{rem}

The corresponding spectral consequence is immediate.

\begin{cor}\label{cor:essential-spectrum}
Let $\Sigma\subset\mathbb R^d$ be the boundary of a bounded Lipschitz domain, let $\alpha\in L^\infty(\Sigma;\mathbb R)$, and let $d\ge2$.
If $F\ne0$, then
$$
\sigma_{\mathrm{ess}}(H_{F,\alpha})=\mathbb R.
$$
Hence
$$
\sigma(H_{F,\alpha})=\mathbb R.
$$
\end{cor}

\begin{proof}
Set
$$
H:=H_{F,\alpha},\qquad H_0:=H_{F,0}.
$$
By Corollary~\ref{cor:selfadjoint-HFa}, the operator $H$ is self--adjoint, and $H_0$ is self--adjoint by construction and Lemma~\ref{lem:free-maximal}.
By Proposition~\ref{prop:compact-resolvent-difference}, the resolvent difference
$$
(H-z)^{-1}-(H_0-z)^{-1}
$$
is compact for $z\in\mathbb C\setminus\mathbb R$.
Hence Weyl's theorem gives
$$
\sigma_{\mathrm{ess}}(H)=\sigma_{\mathrm{ess}}(H_0).
$$
For $F\ne0$,
$$
\sigma_{\mathrm{ess}}(H_0)=\sigma(H_0)=\mathbb R
$$
by Avron--Herbst~\cite{AvronHerbst1977}.
Therefore
$$
\sigma_{\mathrm{ess}}(H)=\mathbb R.
$$
The equality $\sigma(H)=\mathbb R$ then follows from self--adjointness.
\end{proof}

\section{Conclusion}

We treated the Stark Hamiltonian with a real bounded $\delta$ interaction as an elliptic transmission problem on a compact Lipschitz interface.
The transmission class separates the local regularity needed at the interface from the global graph domain of the Stark operator.
This distinction is important because the linear electric potential is unbounded and the free graph domain is not assumed to be contained in $H^1(\mathbb R^d)$.

The direct construction gives a self--adjoint realization and a boundary resolvent formula under the same assumptions.
The boundary equation acts between the natural trace spaces $H^{-1/2}(\Sigma)$ and $H^{1/2}(\Sigma)$.
Compactness of the interface embedding then gives compactness of the resolvent difference.
For a nonzero electric field, the essential spectrum remains equal to $\mathbb R$.
The framework can be used as a starting point for further analysis of scattering and resonances for Stark operators with thin interfaces.

\section*{Statements and Declarations}

\subsection*{Competing Interests}
The author declares that he has no competing interests.

\subsection*{Data Availability}
Data sharing is not applicable to this article as no datasets were generated or analyzed during the current study.

\subsection*{Funding}
The author received no specific funding for this work.

\subsection*{Author Contributions}
The author contributed to all parts of the manuscript.

%%%%%%%%%%%%%%%%

\end{document}